\documentclass[aps,prl,twocolumn,showpacs,superscriptaddress,groupedaddress]{revtex4}  % for review and submission
\usepackage{graphicx}  % needed for figures
\usepackage{dcolumn}   % needed for some tables
\usepackage{bm}        % for math
\usepackage{amssymb}   % for math

\begin{document}

% The following information is for internal review, please remove them for submission
\widetext
%\leftline{Version xx as of \today}
%\leftline{Primary authors: Joe E. Physics}
%\leftline{To be submitted to (PRL, PRD-RC, PRD, PLB; choose one.)}
%\leftline{Comment to {\tt d0-run2eb-nnn@fnal.gov} by xxx, yyy}
%\centerline{\em D\O\ INTERNAL DOCUMENT -- NOT FOR PUBLIC DISTRIBUTION}

% the following line is for submission, including submission to the arXiv!!
%\hspace{5.2in} \mbox{Fermilab-Pub-04/xxx-E}

%\title{Nonlinear traveling waves with oscillatory tails in single column woodpile chains}
\title{Highly Nonlinear Wave Propagation in Elastic Woodpile Periodic Structures}
%\input author_list.tex       % D0 authors (remove the first 3 lines
                             % of this file prior to submission, they
                             % contain a time stamp for the authorlist)
                             % (includes institutions and visitors)
\author{E. Kim}
\affiliation{Aeronautics and Astronautics, University of Washington,
Seattle, WA 98195-2400}

\author{F. Li}
\affiliation{Aeronautics and Astronautics, University of Washington,
Seattle, WA 98195-2400}

\author{C. Chong}
\affiliation{Department of Mathematics and Statistics, University of Massachusetts,
Amherst MA 01003-4515, USA}

\author{G. Theocharis}
\affiliation{Laboratoire d' Acoustique de l' Universit{\'e} du Maine,
UMR-CNRS, 6613 Avenue Olivier Messiaen, 72085 Le Mans, France}

\author{J. Yang}
\affiliation{Aeronautics and Astronautics, University of Washington,
Seattle, WA 98195-2400}

\author{P.G. Kevrekidis}
\affiliation{Department of Mathematics and Statistics, University of Massachusetts,
Amherst MA 01003-4515, USA}

\date{\today}

%\begin{abstract}
%In the present work, we experimentally propose and implement, numerically compute with and
%theoretically analyze a configuration in the form of a 
%single column woodpile periodic structure. One of the theoretically intriguing
%aspects of this chain is its very accurate modeling by
%a Hertzian, locally-resonant, lattice with one or controllably more internal resonators at
%each node. The findings of this study suggest that elastic periodic structures in woodpile configurations
%can be useful not only for manipulating stress waves at will, for example for achieving strong attenuation of high-amplitude impacts, but also for offering a 
%test bed for the formation and study of genuinely traveling waves. In particular, an appealing feature of the woodpile structure is its ability to exhibit a nonlocal traveling wave composed of a 
%strongly-localized solitary wave on top of a small amplitude oscillatory tail. 
%These waves, called {\it sonic-vacuum nanoptera}, are not only motivated theoretically and visualized numerically, but are also 
%observed experimentally by means of a laser Doppler vibrometer. 
%\end{abstract}

\begin{abstract}
In the present work, we experimentally implement, numerically compute with and
theoretically analyze a configuration in the form of a 
single column woodpile periodic structure. %One of the theoretically intriguing aspects of this chain is its very accurate modeling by a Hertzian, locally-resonant, lattice with one or controllably more internal resonators at each node. 
%
%Our findings suggest that a Hertzian, locally-resonant, woodpile lattice
%can be useful for manipulating stress waves at will, for example for achieving strong attenuation and modulation of high-amplitude impacts without relying on damping in the system. Furthermore, we verify that the woodpile architecture offers a test bed for the formation of genuinely traveling waves composed of a 
%strongly-localized solitary wave on top of a small amplitude oscillatory tail. This type of wave, called a {\it nanopteron}, is not only motivated theoretically and numerically, but are also visualized experimentally by means of a laser Doppler vibrometer. 
%
Our main finding is that a Hertzian, locally-resonant, woodpile lattice
offers a test bed for the formation of genuinely traveling waves composed of a 
strongly-localized solitary wave on top of a small amplitude oscillatory tail. This type of wave, called a {\it nanopteron}, is not only motivated theoretically and numerically, but are also visualized experimentally by means of a laser Doppler vibrometer. This system can also
be useful for manipulating stress waves at will, for example, to achieve strong attenuation and modulation of high-amplitude impacts without relying on damping in the system.

\end{abstract}

\maketitle

{\it Introduction.} Granular crystals are rapidly becoming a popular vehicle for the theoretical study, numerical exploration and experimental
identification of a wide range of phenomena ranging from the near linear,
to the weakly or even highly nonlinear limit%. This activity has been summarized in a number of reviews
~\cite{Nesterenko2001,Sen2008,Kevrekidis2011,Theocharis_rev}. The relevant chains consist of assemblies of particles in 
one-, two- and three-dimensions inside a matrix (or a holder) in ordered, 
closely packed configurations. % in which the grains are in contact with each other. 
An especially appealing characteristic of such structures
is the ability to tune their dynamic response by an applied static load. 
This may place the system in a near linear or weakly nonlinear regime, in the case of precompressed chains, or even in a highly nonlinear regime, in the absence of such static load (often termed sonic vacuum, due to the vanishing
sound speed in that case)~\cite{Nesterenko2001}.
It is exactly this dynamic tunability and the controllability of both the assembly
and the measurement of these settings that has enabled a wide range
of proposals for applications. Among others, we note shock and energy absorbing 
layers~\cite{dar06,hong05,doney06}, acoustic lenses \cite{Spadoni}, acoustic 
diodes \cite{Nature11}, and sound scramblers \cite{dar05b}.  
%diodes and switches \cite{Nature11, Li01}, and sound scramblers \cite{dar05b}.  

While various geometries of building blocks have been reported (e.g., spherical, toroidal, or elliptical shapes), granular crystals in woodpile architectures,
made of orthogonally stacked rods, are largely unexplored. This is in contrast to their electromagnetic counterpart -- called woodpile {\it photonic} crystals -- that successfully
demonstrated their efficacy and versatility in manipulating electromagnetic waves ~\cite{Feigel, Liu2007}. Even existing studies on woodpile {\it phononic} crystals
are limited primarily to their linear elastic responses~\cite{Jiang, Wu, Yang2014}, without addressing their nonlinear wave dynamics. 

In this Letter, we show that periodic structures in woodpile configurations can be very useful in manipulating highly nonlinear stress waves at will, including high wave attenuation and spontaneous formation of novel traveling waves after an impact excitation. %Admittedly, the most canonical waveform that arises in nonlinear structures
Arguably, the most fundamental waveform that arises in granular chains within the sonic vacuum is a solitary wave with a highly localized waveform ~\cite{Nesterenko1983,MacKay,pego,Vakakis,atanas,chaterjee}.       %~\cite{Nesterenko1983}.
%
%%  
%\textbf{This particular solution was initially approximated in~\cite{Nesterenko1983}, 
%while a traveling wave was mathematically proved to exist in~\cite{MacKay},
% %and accurately computed numerically in~\cite{pego,Vakakis}, 
% and computed numerically to a prescribed tolerance in~\cite{pego,Vakakis}, 
% while its doubly exponentially decaying form was rigorously proved in~\cite{atanas}. This decay was asymptotically 
%identified originally in the work of~\cite{chaterjee}}.
%
%
%While traveling waves are less common in Hamiltonian 
%lattices~\cite{Kevrekidis2011}
%(with Fermi-Pasta-Ulam type models representing a notable exception),
Recently, other types of coherent traveling waves in granular chains, within the sonic vacuum, were predicted to exist; periodic traveling waves \cite{James1,Vakakis} and static or traveling breathers  in granular chains including on-site potentials~\cite{cuevas}. 

Here, we report experimental evidence of the existence of a new type of nonlocal solitary wave within the sonic vacuum. It consists of a
highly localized solitary waveform on top of an extended, small-amplitude periodic tail, existing in granular chains with local resonators. Such a solution,  satisfying all the other requirements of a solitary wave except that it asymptotes not to zero but to a small amplitude oscillation at infinity, has been long termed a {\it nanopteron}~\cite{boyd_name}.
%~\cite{remois}. %It is for this reason that we will dub the waveform identified herein in the absence of static load a {\it sonic vacuum nanopteron} (SVN). 
This nanopteron arises in
numerous models including continuum~\cite{kruskalseg,buryak1995,akylas} and discrete~\cite{Iooss} dynamical systems.
Some examples, like the $\phi^4$ breather, have received
considerable theoretical attention~\cite{kruskalseg,boyd,luzeng} and relevant
reviews/books have summarized much of this {\it nonlocal
solitary wave} activity~\cite{boyd2,boyd3}. Nevertheless, experimental studies of the nanopteron are
extremely limited~\cite{akylas}.
%~\footnote{An interesting feature in this particular case is that this is based
%on the earlier experimental work of~\cite{farmer} but consists of
%a single figure  that was not published in this work but rather 
%provided to the authors of~\cite{akylas} by Dr. Farmer.}. 

\begin{figure}
\includegraphics[scale=0.4]{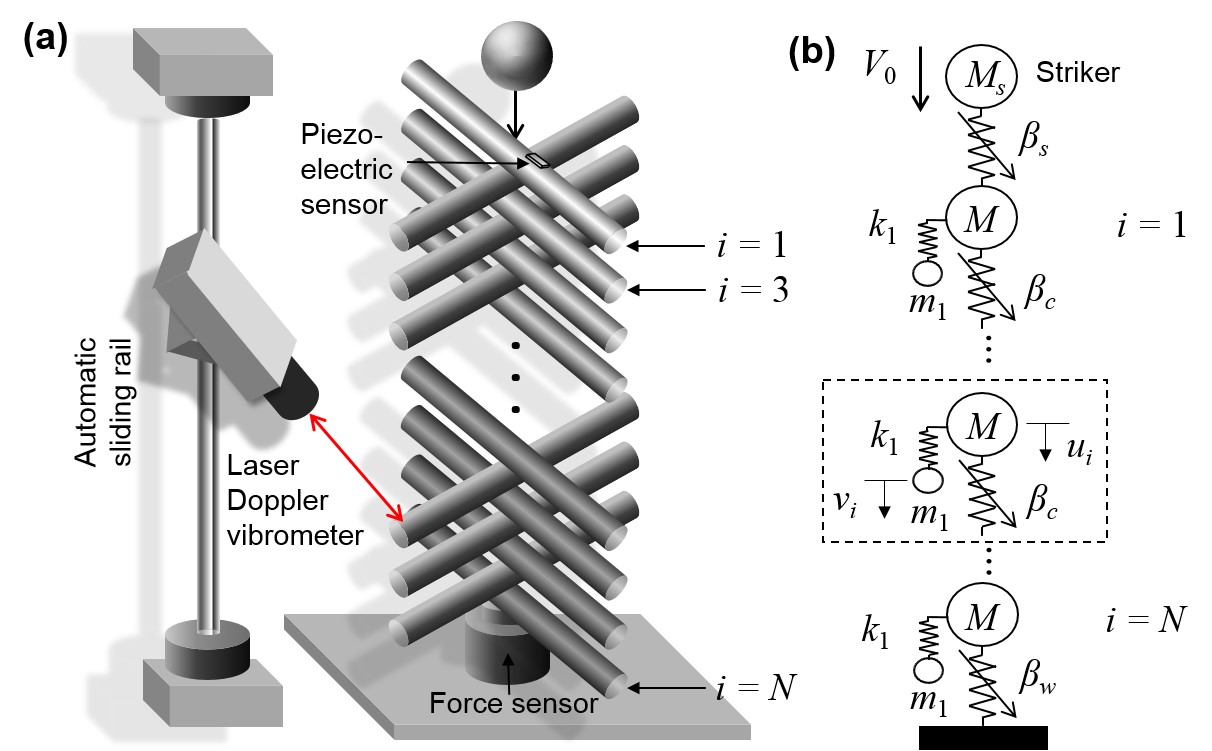}
\caption{Schematic of (a) experimental setup and (b) discrete element model;
see text for details.}
\label{fig_wp1}
\end{figure}

%Thus, one of the goal of the present work is to provide a prototypical example of a nanoptera in the context of the %proposed woodpile structure. In granular lattices, nanopteronic solutions have been reported only in some numerical %simulations in \cite{cuevas}. However, these solutions were observed in a different granular chain model and they have a %different waveform. In our case, the nanopteronic solutions, formed spontaneously by an impact excitation, are highly %nonlinear and consists of a compact compression solitary wave superposed on a small oscillatory tail.
 
%For this, we attempt to obtain for the first time the full-field empirical map of nanoptera by %a non-contact laser Doppler vibrometer.
%example in the context of our proposed woodpile chain (consisting of cylindrical rods)
%of the spontanteous formation of such traveling solitary waves with periodic tails.
%The nanoptera observed are also accurately analyzed by a discrete element model (DEM) bearing %a formal analogy with the granular chain model~\cite{Nesterenko2001},
%but with the addition of an ``internal resonator'' on each node, a
%context that has recently been of considerable interest in its own %right~\cite{serra,luca,gantzounis}. 

In what follows, we present the experimental setup of the
woodpile lattice and a brief overview of its description via an
effective discrete element model (DEM).  
In different regimes,
we experimentally observe (i) the spontaneous formation and steady propagation
of the %SVN
nanopteron, (ii) the potential breathing of the solitary waves, i.e., modulation as they travel
or (iii) the decay of the solitary waves, which is due to the coupling to the
resonators, rather than the damping of the system. All of the relevant features are corroborated by 
numerical computations, and some of the
salient features are explained theoretically. We thus believe that this study
provides a roadmap for further exploration and analysis of highly nonlinear waves 
in a host of settings, including most notably  
granular chain models with the addition of an {\it internal resonator} on each node, a
context that has recently been of considerable interest in its own right~\cite{lazarov,serra,luca,gantzounis,staros2}. 

{\it Experimental and Theoretical Setup.} Figure~\ref{fig_wp1} 
illustrates the experimental setup of 
our 1D woodpile structure and the corresponding DEM. 
The chain is composed of orthogonally stacked cylindrical rods made of fused 
quartz (Young’'s modulus \textit{E} = 72 GPa, Poisson’'s ratio 
\textit{v} = 0.17, and density $ \rho $ = 2200 kg/m$^3$). We test three 
different rod lengths: [20, 40, 80] mm, while keeping their diameters 
identical to 5.0 mm. We excite the chain by striking the center of the 
uppermost rod with a 10 mm-diameter glass sphere. While we present in this Letter the results for a measured impact velocity of 
\textit{V}$_0$ = 1.97 m/s, the effect of varying striker velocities can be found in the supplemental document. 
We record the transmitted stress waves using a 
piezoelectric force sensor (PCB C02) placed at the bottom of the woodpile chain. To investigate the propagating waveforms along the path, we alter 
the number of stacked cylinders from one to \textit{N} (total number of 
cylinders) and synchronize the signals with respect to the striker impact 
moment, which is detected by a small piezoelectric ceramic plate bonded on 
the surface of the top rod. 
%In this study, we also report the first experimental measurements of nanoptera
A particular challenge within our setup concerns the experimental 
identification of the  especially weak oscillations of the unit cells
that are critical for our reported observation of the nanopteron. %SVN. 
For this, we 
introduce a laser Doppler vibrometer (Polytec, OFV-505), which is mounted on an automatic 
sliding rail to detect localized vibrations of each rod in the resolution 
of 0.02 $\mu$m/s/Hz$^{1/2}$.   

As suggested by Fig.~\ref{fig_wp1}(b), the dynamics of the woodpile lattice along the axis of the contacts can be effectively
described by a system of nonlinear oscillators that are coupled
%(directly analogous to the ones used in granular 
%crystals~\cite{Nesterenko2001}) 
to adjacent masses. Assuming the principal nodes (associated with the rods' axial %longitudinal
motion) as having mass $M$ and a coupling of $\beta_c$, and the 
internal resonators within the rods as having a coupling of $k_1$
and a mass of $m_1$, we propose the following generalized Hertzian % discrete element model 
DEM, %can be written in the form
\begin{eqnarray}
M \ddot{u}_i&=&\beta (u_{i-1}-u_i)^{3/2} - \beta (u_i - u_{i+1})^{3/2}
\nonumber
\\
&+& k_1 (v_i - u_i),
\label{eqn1}
\\
m_1 \ddot{v}_i&=&k_1 (u_i - v_i).
\label{eqn2}
\end{eqnarray}
This model allows us to describe longitudinal excitations along the axis of the contacts in the presence of internal vibration modes that can store energy in their own right. 
%From all the internal vibration modes of the rods, due to symmetry reasons only the symmetric bending modes are relevant to the dynamics.
 
The effective parameters $m_1, M$ and $k_1$ of this DEM description are determined via an optimization process based on the envelopes 
of propagating waves (see the supplemental material for further details).
%. However, a key
%additional feature in the present setting is that the rods possess
%internal vibration modes that can store energy in their own right.
Note that in Eq.~(\ref{eqn1}), $\beta$ assumes the value $\beta_c$ within the chain, while
it is $\beta_s$ for the coupling of the striker to the first
bead and $\beta_w$ for the coupling of the last
bead to the wall (cf. Fig.~\ref{fig_wp1}).
In what follows, we will rescale the
time $ t \rightarrow t \sqrt{\beta_c/M}$ and the coupling
$\kappa = k_1/\beta_c$ for the purposes of our analysis. 
%the effective frequency of resonators as $\omega_0 = 
%\sqrt{ \frac{k_1 M}{\beta_c m}}$ and 
The mass ratio is denoted as $\nu=m_1/M$.

\begin{figure}
\includegraphics[scale=0.45]{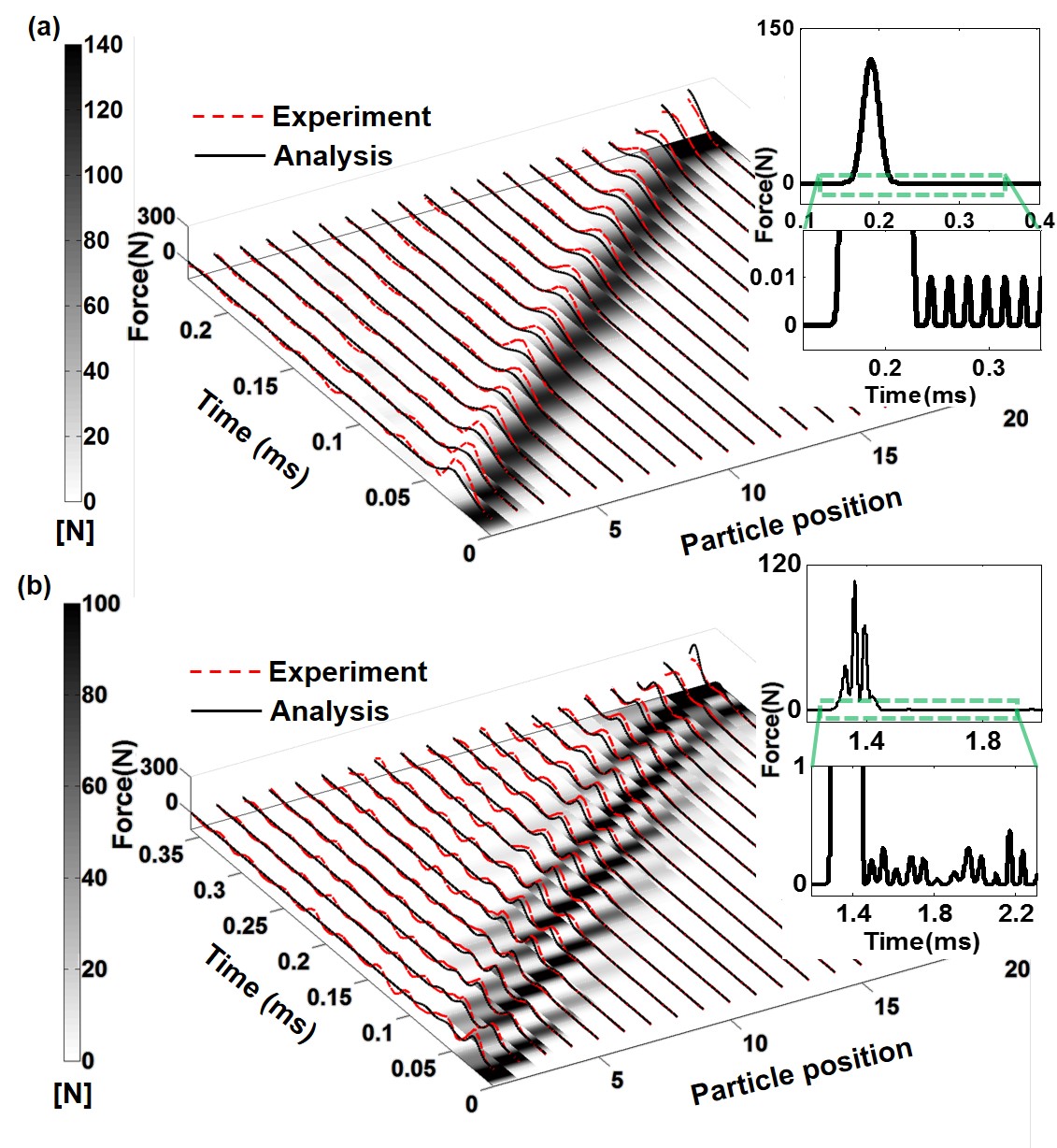}
%\caption{\label{fig:epsart} The first graphs shows wave propagation of 20 mm model. The second graph shows wave propagation of 40 mm model.
%automatically numbered.}
%\end{figure}
%\begin{figure}
%\includegraphics[scale=0.3]{fig2_40mm3.png}
\caption{Numerical (solid black) and experimental (dashed
red) force profiles in space-time (measured in ms) 
in 1D woodpile crystals composed of (a) 20 mm
and (b) 40 mm rods. The insets show the numerical magnified force
profiles of nanoptera, while the colormap represents the magnitude of the contact force.}
%Analytical (solid black) and experimental (dashed red) profiles in
%space-time (measured in ms) of stress waves propagating in 1D 
%woodpile crystals composed of (a) 20 mm and (b) 40 mm rods. The insets show the magnified force profiles of nanoptera, while the colormap represents the magnitude of the contact force.}
\label{fig_wp2}
\end{figure}

{\it Experimental Observations, Numerical Corroboration and Theoretical
Analysis.} Figures~\ref{fig_wp2}(a) and (b) illustrate the comparison
of the wave propagation 
in 1D woodpile lattices %granular crystals 
composed of 20 particles of 20 mm and 40 mm rods respectively. 
Dashed red (solid black) curves represent the contact force profiles obtained by 
experiments (numerics). The numerical results are also shown in the underlying surface maps to ease visualization of wave modulation effects. 
In addition to the accurate representation of the experimental findings
by the DEM, we can make a few further observations here. 
In the case of 20 mm rods, the striker rapidly settles into a solitary wave (in a way
reminiscent of the standard granular chain~\cite{Nesterenko2001,Sen2008} -- 
however with a significant difference, as we will see below). For the
40 mm case, a traveling breather appears to form in a pattern similar to
numerical observations in~\cite{cuevas}.
This  wave emerges after a transient period in which a primary wave experiences an exponential decay
(which can be computed semi-analytically; see supplemental material) and a secondary wave emerges due to the coupling with the resonators.
%~\footnote{These features although of interest in their own
%right merit a separate investigation and will be detailed elsewhere.}. 
However, a key feature shared by {\it both} traveling structures is the
existence of a persistent form of background oscillation as seen in the insets of Fig.~\ref{fig_wp2}. We note that here the wake of the principal
pulse has a constant amplitude tail. 
This feature which has also been confirmed by means of simulations
in considerably larger chains (not shown here) 
is different from what  is the case
in the so-called Kawahara solitary waves, where the tail is 
decaying in amplitude away from the main wave shape~\cite{kawahara}.
We now explore this {\it nanopteronic} waveform more quantitatively.

%the initial traveling wave front experiences an exponential decay of the form $r_i = r_0 c^{i-1}$, where $r_i$ is the %peak
%of the $ith$ lattice site and $c$ is the decay rate.
%he decay rate can be computed semi-analytically as $c\approx 0.8$, see e.g. the supplementary material.
%Due interplay of the primary wave along
%the principal lattice and a secondary wave due to the coupling with the
%esonators, a traveling breather appears to form~\footnote{These features, although of interest in their own
%right, merit a separate investigation and will be detailed elsewhere.}, in a way reminiscent of
%numerical observations in the so-called cradle problem~\cite{cuevas}.
%In the case of
%20 mm rods, the striker rapidly settles into a solitary wave (in a way
%reminiscent of the standard granular chain~\cite{Nesterenko2001,Sen2008} -- 
%however with a significant difference, as we will see below). 

%A key feature shared by {\it both} traveling structures shown in Fig.~\ref{fig_wp2}(a) and (b) is the
%existence of a persistent form of background oscillation, % that we now explore more quantitatively.
%The most interesting feature that, arguably, both waveforms
%of Fig.~\ref{fig_wp2} share is the existence of a backdrop of
%apparent small amplitude oscillations on the wake of the principal
%solitary wave.
%These modulations 
%which is analyzed in further detail in Fig.~\ref{fig_wp3}.

\begin{figure}
\includegraphics[scale=0.6]{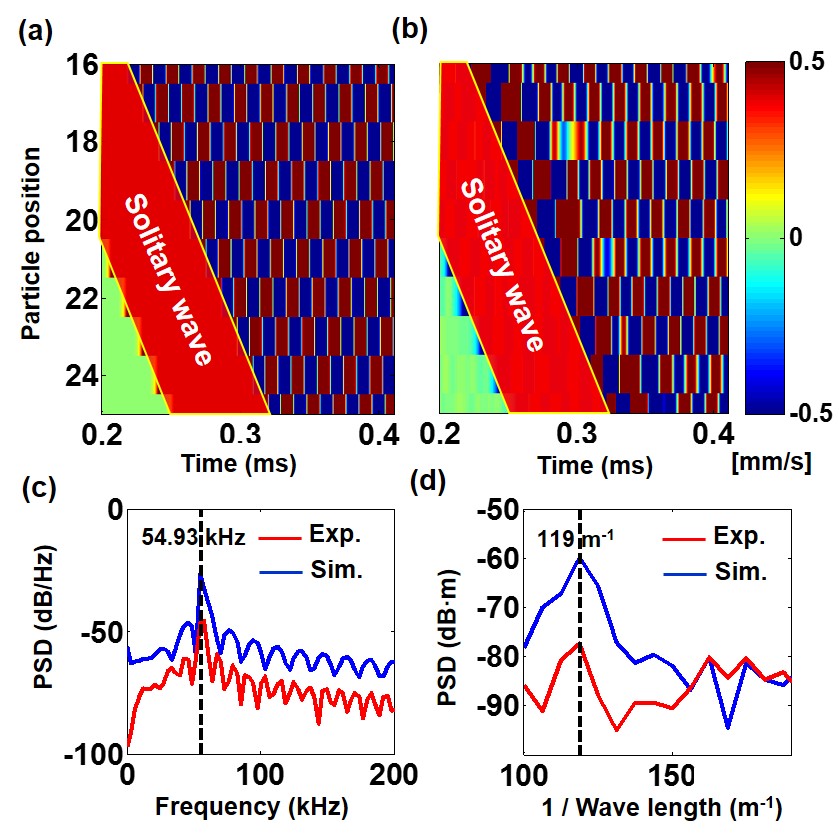}
\caption{(a) Numerical and (b) experimental velocity profiles of nanoptera formed in a 40 particle chain of 20 mm rods. (c) Frequency and (d) wave number contents of the tail constructed
by FFT of velocity profiles in specific time- and space-domains, respectively (particle spot $i$ = 24 and time $t$ =  0.4 ms).}
\label{fig_wp3}
\end{figure}

The surface maps in Fig.~\ref{fig_wp3}(a) and (b) show the analytical and 
experimental velocity profiles respectively of the tails of the observed
waveforms that appear in a 40 particle chain of 20 mm rods. The traveling
waves 
spontaneously become nanoptera %SVN 
by developing oscillatory patterns of velocity,
which clearly follow 
the principal solitary wave (highlighted in red color). It should be noted 
that the velocities involved in 
the nanopteronic tails are approximately three orders 
of magnitude smaller than those of the solitary waves; yet,  
they can be accurately measured through our laser Doppler vibrometer. The frequency and 
wavenumber content of the nanopteronic % SVN 
tail can be obtained by conducting the fast 
Fourier transform (FFT) in time- and space-domains (shown in
Fig.~\ref{fig_wp3}(c) and (d)). The resonant frequency of the experimental data shown in 
panel (c) is 54.93 kHz, which is found to be directly connected to the relative
motion of the two masses (the primary and the resonator ones),
namely $\omega_0=\sqrt{\kappa (1+1/\nu)}$ (55.45 kHz according to the DEM). For a traveling wave of speed
$c$, the corresponding wavenumber in panel (d) 
is found to satisfy the relation $\omega_0=c k_0$. In Fig.~\ref{fig_wp3}(d), we obtain $k_0=119$ m$^{-1}$ experimentally, which is in agreement with the value $k_0=120$ m$^{-1}$ obtained via the DEM (see the supplemental document for details).

%{\bf JK: we should probably give some specifics to corroborate
%this claim here, namely give the $\nu$, $\kappa$, $c$ and connect
%with the observed 55.2 KHz and the $k_0=119$ m$^{-1}$}.
%ion also is presented in Fig 4 (d). Dashed line is the wave number 
%calculated from the relation 

We now theoretically justify this feature, namely the existence of the relative motion between 
the primary node and the resonator, in the nanopteronic tail of the observed wave structure. %SVN.
Setting up the so-called strains of the two fields
$r_i=u_{i-1}-u_i$ and $s_i=v_{i-1}-v_i$, seeking traveling waves
therein as $r_i(t)=R(i-c t)=R(\xi)$, $s_i=S(i-c t)=S(\xi)$ and then
using the Fourier transform 
$R(\xi)=\int_{-\infty}^{\infty} \hat{R}(k) e^{i k \xi} dk$ (and similarly
for $S$), leads from Eqs.~(\ref{eqn1})-(\ref{eqn2}) to
\begin{eqnarray}
\hat{R} &=& \frac{1}{c^2} {\rm sinc}^2\left(\frac{k}{2}\right)\widehat{R^{3/2}} + \frac{\kappa}{k^2c^2}(\hat{R}-\hat{S}),
\label{eqn3_mim}
\\
\hat{S} &=& \frac{\kappa}{\kappa-c^2 k^2 \nu} \hat{R}.
\label{eqn4_mim}
\end{eqnarray}
% \begin{eqnarray}
% \hat{R} &=& \left( \frac{1-c^2 k^2 \nu}{1+\nu - c^2 k^2 \nu} \right) 
% \frac{1}{c^2} {\rm sinc}^2\left(\frac{k}{2}\right) \widehat{R^{3/2}}
% \label{eqn3_mim}
% \\
% \hat{S} &=& \frac{1}{1-c^2 k^2 \nu} \hat{R}
% \label{eqn4_mim}
% \end{eqnarray}
Substituting Eq.~(\ref{eqn4_mim}) into Eq.~(\ref{eqn3_mim})
and reshaping the relevant expression yields
%using $k_n=
%(1+ 1/\nu)/c^2$, we have:
\begin{eqnarray}
\hat{R}=\left[\frac{1}{c^2} {\rm sinc}^2\left(\frac{k}{2}\right)  + \frac{1}{c^4} \frac{\kappa}{k^2-k_0^2} {\rm sinc}^2\left(\frac{k}{2}\right) \right]
\widehat{R^{3/2}}.
\label{eqn5_mim}
\end{eqnarray}
Recall that sinc$(x)=\sin(x)/x$.
%Using properties of the convolution (assuming $c=1$ for simplicity without
%loss of generality) allows us to write
Invoking the convolution theorem 
%(assuming $c=1$ for simplicity without
%loss of generality, and that the technical assumptions for the theorem hold), 
leads us to write
\begin{eqnarray}
R(x) = K \ast R^{3/2} = \int_{-\infty}^{\infty} K(x-y) R^{3/2}(y) dy,
\label{eqn6_mim}
\end{eqnarray}
where $K(x)= \Lambda(x) + M(x)$, where $\Lambda(x)= 
(1/c^2) \max(1-|x|,0)$ and appears in the corresponding calculation for the
granular chain without internal resonators~\cite{pego}. For $M(x)$ we find
\begin{eqnarray}
&&(2 c^4 k_0^3/\kappa) M(x) = |1-x|
({\rm sinc}(k_0 (1-x))- k_0 )  \label{eqn7_mim}  \\ && -2 |x| ({\rm sinc}(k_0 x)-k_0)
%\nonumber 
+ |x+1| ({\rm sinc}(k_0 (x+1))-k_0). 
%\nonumber
%\\
\nonumber
\end{eqnarray}
Thus, the sinusoidal dependence
with the periodicity dictated by $k_0$ within $M(x)$ is directly
responsible for the formation of the nanopteronic tails;
cf. also the resonant term in the Fourier space
expression of Eq.~(\ref{eqn5_mim}). In the
granular chain without the resonators, the presence of solely the
$\Lambda$ term in Eq.~(\ref{eqn5_mim}) produces a monotonically
decaying solitary wave according to a double exponential law~\cite{pego,atanas}.
%We should note here that direct iterations of Eq.~(\ref{eqn5_mim})
%allows one to construct the numerically exact form of the nanopteron solution
%(results not shown here), which, as shown, is supported against the backdrop
Here, the presence of the sinusoidal terms within $M(x)$ justifies the form of the nanopteron, %SVN,
where the localized central wave is supported against the backdrop
of linear relative vibrations between each node and its corresponding
resonator.

\begin{figure}
\includegraphics[scale=0.45]{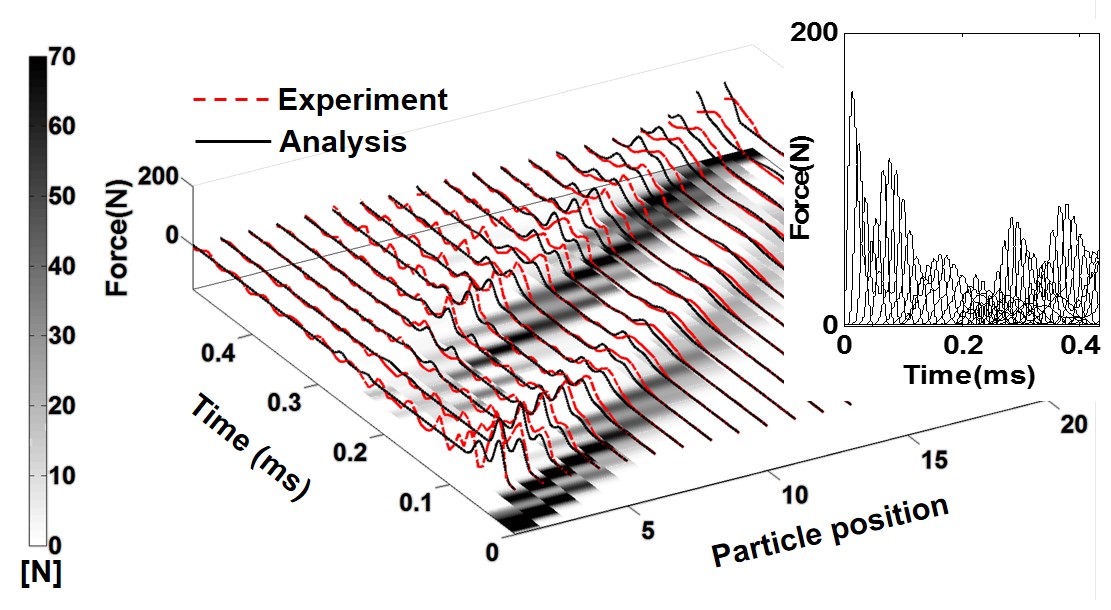}
\caption{ Experimental and numerical space-time wave modulation 
results in woodpile woodpile crystals composed of 80 mm rods.}
\label{fig_wp4}
\end{figure}

Finally, it should be noted that the present setup provides
numerous additional opportunities for a wide range of studies
within this class of models. 
One such consists of modifying the rod length. For example, the
experimental and numerical results
for $80$ mm rods is shown in Fig.~\ref{fig_wp4}. In this case, 
the DEM needs to account for two internal resonant modes within
the rod and hence two resonators $(v_i,w_i)$ are attached to each principal
node of the chain ($u_i$). As a result, we observe that in this
case, 
%it is no longer straightforward for either a solitary
%wave to be formed or for a breathing structure to emerge. Instead,
the large-amplitude striker impact drastically decays through an effective excitation of the internal resonant modes which disperse the energy in both
the temporal and the spatial domain. The inset of Fig.~\ref{fig_wp4} depicts the overlapped profiles of nonlinear waves obtained from various particle positions, which evidently indicate the decaying trend of the propagating waves due to the coupling to the resonators. This wave attenuation suggests that the woodpile periodic structure could be used as an efficient impact mitigator without relying on damping in the system. We should note here that although in this exposition we have highlighted
some of the salient features of the model, numerous additional
details including the experimental setup, the precise selection of the
DEM parameters and the quantitative nature of the agreement between
theory, numerics and experiment are provided
in the supplemental material (see e.g. Fig. 8 therein). %Supplemental Material.

%, although a secondary, weaker structure
%appears to form and persist at least for the chain lengths
%considered in our experiments and numerical simulations. 
%In addition to the possibility to
%tune the resonant frequencies and coupling strengths within
%each unit cell $(u_i,v_i)$

{\it Conclusions and Future Challenges.} In the present work, we have 
offered a prototypical example of a woodpile granular crystal, consisting of
a chain of orthogonally stacked cylindrical rods. In addition
to developing the experimental techniques enabling a distributed
space-time sensing of the chain, we have provided a theoretical discrete element
 model that captures the fundamental experimental
characteristics of the system, while generalizing the standard
Hertzian chain via the inclusion of at least one or modularly
more on-site resonators. We have seen that this inclusion 
provides the possibility for a potential breathing traveling wave or even
decay of the initial strong impulse. More importantly, the
relative motion between each node and the attached resonator
provides the linear mode which constitutes the background for
the formation of a weakly
nonlocal solitary wave, i.e., a %sonic-vacuum 
nanopteron. Despite the
small magnitude of the tails of the nanoptera (differing by three orders
of magnitude with respect to the principal wave), we were able
to experimentally observe and compute these tails and to theoretically account
for the wavenumber/frequency of their periodicity.

This study leads to a number of topics %questions 
for potential future work.
From a rigorous mathematical perspective, proving the existence of the nanopteron
% (e.g., at the level of Eq.~(\ref{eqn6_mim})) 
provides a novel
set of challenges. At the discrete element model level, quantifying
the properties of the system in the case of one or more resonators by detailing
the interplay between principal and secondary waves or the role
of parametric variations (such as tuning the resonant frequency of the 
coupling between unit cells etc.) would be of particular interest.
It is also relevant to point out that our numerically/experimentally
observed nanoptera have a tail only on one side (i.e., are ``one-sided''
nanoptera), while the typical examples previously known have tails
on both sides. Understanding when one-sided vs. two-sided installments
of such coherent structures may arise could be of particular interest
for future work. In the same vein, considering the results of collisions
of two such (e.g. counter-propagating) waves could also shed
light on the robustness of such one-sided nanoptera, as well as
potentially lead to the formation of two-sided variants thereof.
Finally, several questions naturally emerge in experimental investigations.
This includes examining the problem in the presence of precompression
and its generalization to higher order settings. %which would provide
%valuable insights towards future possibilities in the theory or applications
%of the woodpile granular crystals.
From a practical perspective, this woodpile structure 
can offer a new way to modulate, localize, or mitigate external impacts for 
engineering devices and associated applications.

{\it Acknowledgements.} We thank C. Daraio, D. Khatri and D. Frantzeskakis for helpful discussions. JK and PGK 
acknowledge the support of NSF (CMMI-1414748, CMMI-1000337, DMS-1312856), AFOSR (FA9550-12-1-0332) and ONR (N000141410388).
GT aknowledges financial support from FP7-CIG (Project 618322 ComGranSol).


\begin{thebibliography}{99}

\bibitem{Nesterenko2001} V.F. Nesterenko,  Dynamics of Heterogeneous Materials, Chapter 1, Springer-Verlag (New York, 2001).

\bibitem{Sen2008}	S. Sen, J. Hong, J. Bang,  E. Avalosa, R. Doney,  
%Solitary waves in the granular chain. 
Physics Reports {\bf 462}, 21-66 (2008).

\bibitem{Kevrekidis2011} P.G. Kevrekidis,
IMA J. Appl. Math. {\bf 76}, 389 (2011).

\bibitem{Theocharis_rev} G. Theocharis, N. Boechler, and C. Daraio,
% “Nonlinear Phononic Periodic Structures and Granular Crystals”,
in {\it Phononic Crystals and Metamaterials}, Ch. 6, 
Springer Verlag,
(New York, 2013)


\bibitem{dar06} C. Daraio, V.~F. Nesterenko, E.~B. Herbold, and S. Jin, Phys. Rev. Lett. {\bf 96}, 058002 (2006).

\bibitem{hong05} J. Hong, Phys. Rev. Lett. {\bf 94}, 108001 (2005).

\bibitem{doney06} R. Doney and S. Sen, Phys. Rev. Lett. {\bf 97}, 155502 (2006).

\bibitem{Spadoni} A. Spadoni and C. Daraio, Proc Natl Acad Sci USA, {\bf 107}, 7230, (2010).


\bibitem{Nature11} N. Boechler, G. Theocharis, C. Daraio,  Nature Materials \textbf{10}, 665 (2011). %-668.
%\emph{Bifurcation based acoustic switching and rectification},

%\bibitem{Li01} F. Li, P. Anzel, J. Yang, P.G. Kevrekidis, and C. Daraio, 
%Nat. Comm. {\b 5}, 5311 (2014).

\bibitem{dar05b} V.~F. Nesterenko, C. Daraio, E.~B. Herbold, and S. Jin, Phys. Rev. Lett. {\bf 95}, 158702 (2005).


\bibitem{Feigel} A. Feigel et al., 
Appl. Phys. Lett. {\bf 83}, 4480 (2003).

\bibitem{Liu2007} H. Liu, J. Yao, D. Xu, and P. Wang, 
Optics Express. {\bf 15}:2, 695 (2007).

\bibitem{Jiang} H. Jiang, Y. Wang, M. Zhang, Y. Hu, D. Lan, Y. Zhang, and B. Wei,
Appl. Phys. Lett. {\bf 95}, 104101 (2009).

\bibitem{Wu} L.Y. Wu and L.W. Chen,
J. Phys. D: Appl. Phys. {\bf 44}, 045402 (2011).

\bibitem{Yang2014} E. Kim and J. Yang,
J. Mech. Phys. Solids {\bf 71}, 33 (2014).

\bibitem{Nesterenko1983} V.F. Nesterenko,  
%\emph{Propagation of nonlinear compression pulses in granular media},
 J. Appl. Mech. Techn. Phys.  {\bf 24}, 733 (1983).


\bibitem{MacKay} R.S.~MacKay, Phys. Lett. A {\bf 251}, 191 (1999).

\bibitem{pego} J.~M. English and R.~L. Pego,
Proceedings of the AMS {\bf 133}, 1763 (2005).

\bibitem{Vakakis} Y.~Starosvetsky and A.F.~Vakakis, Phys. Rev. E {\bf 82}, 026603 (2010).

\bibitem{atanas} A.~Stefanov and P.~Kevrekidis,
%\newblock \doititle{Traveling waves for monomer chains with precompression},
\newblock {Nonlinearity} \textbf{26}, 539 (2013).

\bibitem{chaterjee} A. Chaterjee, Phys. Rev. E {\bf 59}, 5912 (1999).

\bibitem{James1} G. James, J. Nonlinear Sci. {\bf 22} 813 (2012).

\bibitem{cuevas} G. James, P.G. Kevrekidis and J. Cuevas,
Phys. D {\bf 251}, 39 (2013).

\bibitem{boyd_name} J.P. Boyd, 
``Weakly non-local solitary waves" pp. 51-97 in 
Nonlinear Topics in Ocean Physics: Proceedings of the Fermi School, 
%edited by 
A.R. Osborne and L. Bergamasco Eds. (Amsterdam: North-Holland, 1991)

%\bibitem{remois} M. Remoissenet, {\it Waves called solitons},
%Springer-Verlag (Berlin, 1999); see, in particular, pp. 176-177.

%\bibitem{zolotaryuk1998} 

\bibitem{kruskalseg} M. Kruskal, H. Segur, 
Phys. Rev. Lett. {\bf 58}, 747 (1987).


\bibitem{buryak1995} A.V. Buryak, Phys. Rev. E {\bf 52}, 1156 (1995);
N.N. Akhmediev, A.V. Buryak, Opt. Commun. {\bf 121}, 109 (1995).

\bibitem{akylas} T. Akylas and R. Grimshaw,
J. Fluid Mech. {\bf 242}, 279 (1992). An interesting feature in this particular case is 
that it consists of a single figure based on the earlier experimental work of D. Farmer and J. Smith,
Deep-Sea Res. 27. 239 (1980), which was provided to T. Akylas and R. Grimshaw by Dr. Farmer.

\bibitem{Iooss} G. Iooss, G.~James, Chaos {\bf 15}, 015113 (2005).

\bibitem{boyd} J.P. Boyd, 
Nonlinearity {\bf 3}, 177 (1990).


\bibitem{luzeng} N. Lu,
J. Diff. Eqs. {\bf 256}, 745 (2014).


\bibitem{boyd2} J. Boyd, {\it Weakly nonlocal solitary waves
and beyond-all-orders asymptotics}, Kluwer (Amsterdam, 1998).

\bibitem{boyd3}  J.P. Boyd,
Acta Applicandae {\bf 56}, 1 (1999).




%\bibitem{farmer} D. Farmer and J. Smith,
%Deep-Sea Res. {\bf 27}, 239 (1980).

\bibitem{lazarov} B.S. Lazarov, J.S. Jensen, 
%Low-frequency band gaps in chains with attached non-linear oscillators, International Journal of Non-Linear Mechanics 42 
J. Nonlin. Mech. {\bf 42}, 1186 (2007).

\bibitem{serra} P.G. Kevrekidis, A. Vainchtein, M. Serra-Garcia,
Phys. Rev. E {\bf 87}, 042911 (2013).

\bibitem{luca} L. Bonanomi, G. Theocharis, C. Daraio,
 arXiv:1403.1052.

\bibitem{gantzounis} G. Gantzounis, M. Serra-Garcia, K. Homma,
J.M. Mendoza, C. Daraio, J. Appl. Phys. {\bf 114}, 093514 (2013).


\bibitem{staros2} K. Vorotnikov, Y. Starosvetsky,
G. Theocharis, P.G. Kevrekidis,
preprint, submitted to Phys. D (2013).

\bibitem{kawahara} T. Kawahara,
J. Phys. Soc. Jpn {\bf 33}, 260 (1972); see also:
C.I. Christov, G.A. Maugin, M.G. Velarde,
Phys. Rev. E {\bf 54}, 3621 (1996).

\bibitem{prl110} E. B. Herbold and V. F. Nesterenko,
Phys. Rev. Lett. {\bf 110}, 144101 (2013).

%  \bibitem{d0det}
%    Standard D\O\ detector reference:  \\
%V.M.~Abazov {\sl et al.} (D0 Collaboration),
%Nucl. Instrum. Methods Phys. Res. A {\bf 565}, 463  (2006).

%  \bibitem{d0lumi}
%    ** New ** D\O\ luminosity reference: \\
%T.~Andeen {\sl et al.}, FERMILAB-TM-2365 (2007).

%  \bibitem{pdg}
%   Particle Data Group reference: \\
%   W.-M.~Yao {\sl et al.}, Journal of Physics G {\bf 33}, 1 (2006).

%  \bibitem{geant}
%    {\sc Geant} reference: \\
%    R. Brun and F. Carminati, CERN Program Library Long Writeup W5013, 1993 (unpublished).

%  \bibitem{pythia}
%    {\sc Pythia} reference: \\
%    T. Sj\"{o}strand {\sl et al.}, Comput. Phys. Commun. {\bf 135}, 238 (2001).

%  \bibitem{cteq}
%    {\sc Cteq6} reference: \\
%    J. Pumplin {\sl et al.}, JHEP {\bf 0207} 012 (2002) and
%    D. Stump {\sl et al.}, JHEP {\bf 0310} 046 (2003).

%  \bibitem{cls}
%    LEP CL$_S$ reference: \\
%    T. Junk, Nucl. Instrum. Methods A {\bf 434}, 435 (1999).

%  \bibitem{d0limit}
%   D\O\ Bayesian reference: \\
%   I. Bertram {\sl et al.}, FERMILAB-TM-2104 (2000).

%  \bibitem{d0jets}
%  D\O\ cone-jet reference: \\
%G.C.~Blazey {\it et al.}, in
%{\sl Proceedings of the Workshop: QCD and Weak Boson
%Physics in Run II,} edited by U.~Baur, R.K.~Ellis, and
%D. Zeppenfeld, Fermilab-Pub-00/297 (2000).
%
\end{thebibliography}
\end{document}

% The following information is for internal review, please remove them for submission
\widetext
%\leftline{Version xx as of \today}
%\leftline{Primary authors: Joe E. Physics}
%\leftline{To be submitted to (PRL, PRD-RC, PRD, PLB; choose one.)}
%\leftline{Comment to {\tt d0-run2eb-nnn@fnal.gov} by xxx, yyy}
%\centerline{\em D\O\ INTERNAL DOCUMENT -- NOT FOR PUBLIC DISTRIBUTION}

% the following line is for submission, including submission to the arXiv!!
%\hspace{5.2in} \mbox{Fermilab-Pub-04/xxx-E}

\title{SUPPLEMENTAL MATERIAL:\\
Highly Nonlinear Wave Propagation in Elastic Woodpile Periodic Structures}
%Nanopteronic Highly Nonlinear Waves in Woodpile Granular Crystals:\\
% Theory, Computation and Experiment}

\author{E. Kim}
\affiliation{Aeronautics and Astronautics, University of Washington,
Seattle, WA 98195-2400}

\author{F. Li}
\affiliation{Aeronautics and Astronautics, University of Washington,
Seattle, WA 98195-2400}

\author{C. Chong}
\affiliation{Department of Mathematics and Statistics, University of Massachusetts,
Amherst MA 01003-4515, USA}

\author{G. Theocharis}
\affiliation{Laboratoire d' Acoustique de l' Universit{\'e} du Maine,
UMR-CNRS, 6613 Avenue Olivier Messiaen, 72085 Le Mans, France}

\author{J. Yang}
\affiliation{Aeronautics and Astronautics, University of Washington,
Seattle, WA 98195-2400}

\author{P.G. Kevrekidis}
\affiliation{Department of Mathematics and Statistics, University of Massachusetts,
Amherst MA 01003-4515, USA}

\pacs{45.70.-n 05.45.-a 46.40.Cd}
\date{\today}

\maketitle

%\paragraph{Experimental setup.} 
\section{Experimental setup.} 
\begin{figure}
\includegraphics[scale=0.5]{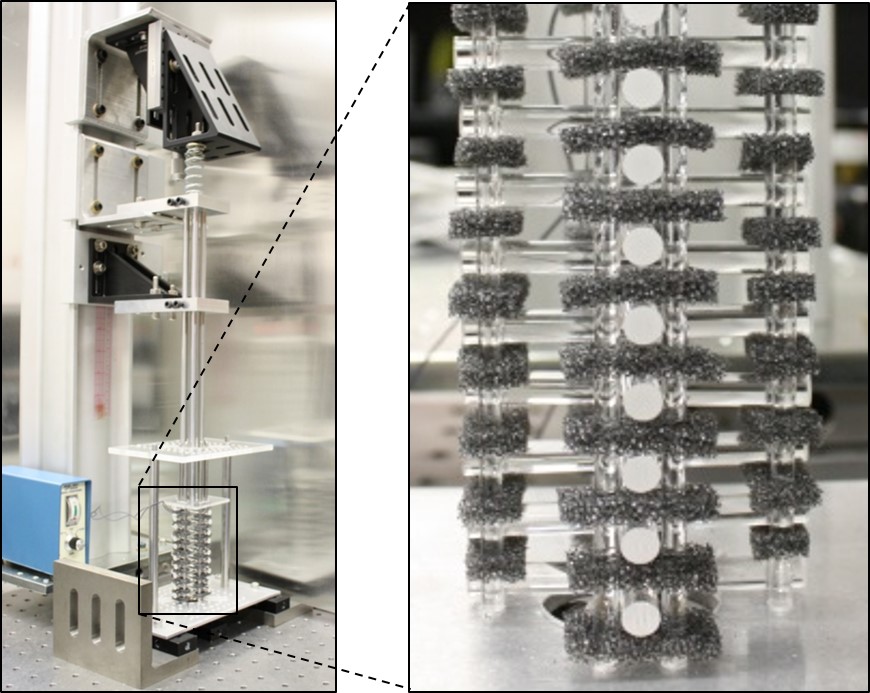}
\caption{ Experimental setup of a woodpile granular crystal composed of orthogonally stacked 40 mm rods. A magnified view is shown on the right.}
\label{ExpSetup}
\end{figure}

Figure \ref{ExpSetup} shows digital images of the experimental woodpile setup constructed in this study.
 The chain is supported by vertically-standing guiding rods, which constrain the horizontal, in-plane motions of the cylindrical 
elements and prevent buckling of the chain. To align the rods along the chain in parallel, we position soft polyurethane foam 
(firmness: 4.1 - 6.2 kPa under 25\% deflection) between the tips of the rods. 
We find that the effect of the supporting foam on the cylinders' motions during wave propagation is negligible, 
because of the orders-of-magnitude lower stiffness of the foams compared to the rigidity of the cylinders or the 
contact stiffness of the vertically stacked cylindrical rods. 
As a result, we clearly observe the nanopteron tails with the Laser Doppler vibrometer 
%despite that the magnitude of the vibrational veolocity is extremely small in 
despite that the amplitude of the vibration is extremely small in 
the order of  tens of nm. 
%{\bf We measure nanopteron tails in the chain having 40 rods with 20 mm length, while the force profile is measured based on the chain having 20 rods in each model.}
\\

\begin{figure}
\includegraphics[scale=0.45]{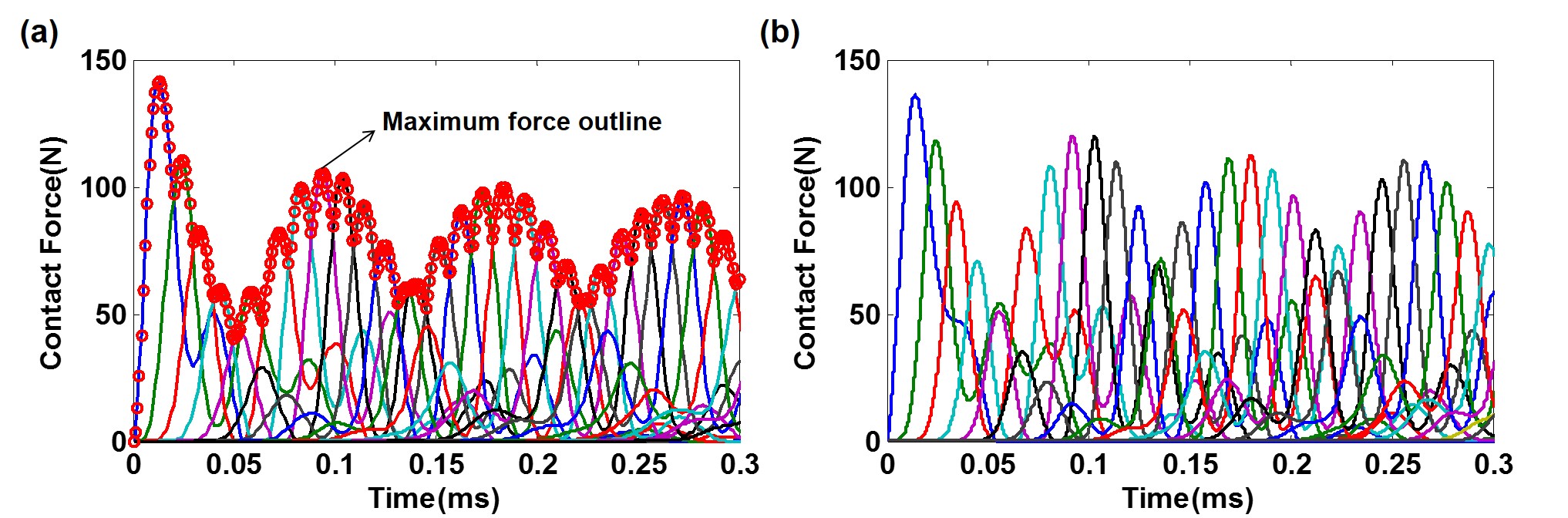}
\caption{Contact force profiles for a 40 mm woodpile granular crystal from (a) FEM simulation and (b) DEM simulation
 with optimized parameters. }
\label{optimization}
\end{figure}

\begin{table*}
\label{tab:param}
\caption{Discretized mass distribution and spring coefficients in woodpile granular crystals.}
\begin{ruledtabular}
\begin{tabular}{ccccccc}
Rod length & Total mass ($g$) & $M$ ($g$) & $m_1$ ($g$) &$m_2$ ($g$) &$k_1$ ($kN/m$)  &$k_2$ ($kN/m$) \\
\hline
20 mm & 0.866 & 0.423 & 0.443 & - & 26267 & - \\
40 mm & 1.732 & 0.838 & 0.894 & - & 3961&- \\
80 mm & 3.464 & 1.00 & 1.80 & 0.664 & 532 & 6423 \\
\end{tabular}
\end{ruledtabular}
\end{table*}

%\paragraph{Parameter determination for DEM.}
\section{Parameter determination for DEM.}
For the construction of the DEM, we need to determine the parameters of a unit cell in terms of its discretized mass and 
spring coefficients (e.g., $M$, $m_1$, and $ k_1$ as shown in the boxed region of Fig 1 (b)). In this study, 
these parameters are calculated via an optimization process based on the finite element method (FEM) (details are described in the reference [15] of the main manuscript).
 For example, Fig.~\ref{optimization}(a) shows the temporal profiles of numerically simulated contact forces in the case of a 
40 mm woodpile granular crystal. From the FEM result, it is evident that this locally resonating structure develops a modulated 
shape of propagating waves, which is in agreement with the experimental results as presented in Fig.~2(b). For the optimization 
process, we extract the overall shape of the modulated waveform by taking the maximum values of overlapped contact force 
 (see red markers in Fig.~\ref{optimization}(a)). The next step involves tuning the parameters of the DEM to match this 
baseline curve obtained by the FEM. For this, we begin with an initial mass distribution, $M$ and $m_1$, and compute 
the stiffness of $k_1$ based on $\omega_0=\sqrt{\kappa (1 + 1/\nu)}$, where $\kappa = k_1/\beta_c$ and $\nu=m/M$.
 Note that $\omega_0$ is a resonant frequency of the unit cell (i.e., 40 mm rod) calculated from the FEM, and $\beta_c$ 
is given simply by the Hertzian contact law. Based on these mass and stiffness values, we simulate the wave propagation
 via the DEM and obtain the modulated waveforms as shown in Fig.~\ref{optimization}(b). We then compare it with the baseline
 data and iterate this process until we get optimized parameters that yield the best match between the FEM and the DEM. 
The optimized parameters for discretized masses and spring coefficients are summarized in Table I for woodpile granular
 chains with rod lengths of 20, 40, and 80 mm. Note, from all the possible internal vibration modes of the rods, 
 only the symmetric bending modes are relevant to the dynamics due to symmetry considerations~\cite{Yang2014}.
\\

\begin{figure}
\includegraphics[scale=0.45]{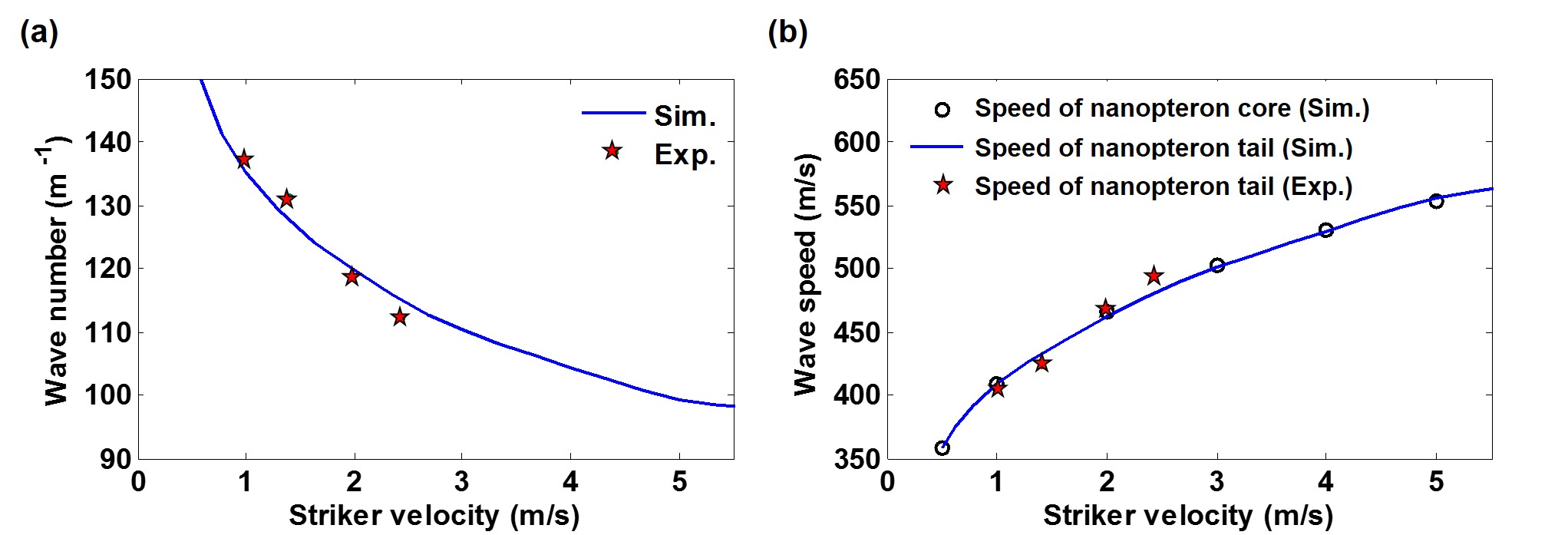}
\caption{ (a) Wavenumber and (b) speed of sonic vacuum nanopteronic (SVN) tails (i.e., wings) as a function of various striker velocities (20 mm rods).
 Blue curves denote DEM results, while the start markers are experiment data measured from four different striker velocities: [0.99 1.40 1.97 2.42] m/s. 
%while the star markers correspond to the experimental verifications given the testing data as presented in Fig.~3(c) and (d). 
The wave speed is also calculated alternatively by tracing the solitary waves 
(i.e., core part of the nanopteron) as marked by hollow circles.}
\label{strikerVelocity}
\end{figure}

%\paragraph{Effect of striker velocity variation.} 
\section{Effect of striker velocity variation.} 
The speed of solitary waves in a discrete chain connected with Hertzian-type contact depends on the wave amplitude,
 i.e., $c \propto F_m^{1/6}$ where $F_m$ is the force magnitude of propagating waves. This force magnitude is 
strongly dependent on striker velocity~\cite{coste01}. Thus, it can be interpreted that the speed of nanopteron (i.e., both of the solitary wave part and tail) 
%Thus, it can be interpreted that the speed of the solitary wave part of the nanoptera (i.e., their core) 
increases as we impose a higher 
striker velocity on the woodpile granular crystal. This is because the wave speed of the nanopteronic %SVN 
tail is the same as the speed of the core, 
namely $c=\omega_0/k_0$, where $\omega_0$ and $k_0$ are the frequency and wavenumber of the nanopteronic, %SVN tail,
 respectively. We numerically verify this using the chain composed of 20 mm rods under various striker 
velocities (Fig.~\ref{strikerVelocity}). Specifically, we first compute velocity profiles of all particles 
in the chain using the DEM. Based on the velocity profiles of the nanopteronic %SVN 
tails, we obtain wavenumbers 
and frequencies by conducting FFT in the space- and time-domain, respectively. The wavenumber varies as a 
function of striker velocity as shown in Fig.~\ref{strikerVelocity}(a), while the extracted frequencies 
coincide with the natural frequency of the cylindrical elements. The wave speed is obtained directly 
from $\omega_0/k_0$ (blue curve in Fig.~\ref{strikerVelocity}(b)). This is in agreement with the speed 
of the core part of the nanopteron (i.e., solitary waves' speed) as marked by hollow circles in 
Fig.~\ref{strikerVelocity}(b). Experimental results corroborate the numerical simulations by 
the DEM (star markers in Fig.~\ref{strikerVelocity}).  

\section{Semi-analytical description of wave propagation}

In the absence of the local resonators,  Eq.~(1) possesses exact traveling wave solutions \cite{friesecke94,pego05,stefanov1}. Two classical
approaches to obtain an analytical approximation to the
evolution of the traveling wave in the presence of perturbations are
the map approach~\cite{staros}
%the long wavelength approximation 
%\cite{Nesterenko2001} 
and the binary collision approximation (BCA) \cite{Lindenberg}. 
In this section, we amend these two methods in order to study
the traveling waves in the presence of the local resonators. 
It should be noted here that the map approach presented
below in the presence of
resonators was already developed in the work of~\cite{staros2},
whose discussion we adapt for our present comparison with 
the experimental results.

\subsection{A Map Approach}

%We first consider an alternative method to approximate the dynamics of the discrete element model~Eq.~\eqref{eq:5}
%that will make use of Green's functions and the long wavelength approximation of the traveling wave in the 
%Using the rescaling,
%$$  t \rightarrow t \, \sqrt{\frac{\beta_c}{M}}, \qquad \kappa = \frac{k_1}{beta_c}, \qquad \nu = \sqrt{ \frac{m}{M} },   $$
Ignoring the boundary effects (i.e. considering an infinite chain), we write the equations of motion in terms of the strain variables $r_{i} = u_{i-1}-u_i$
and $s_i = v_{i-1} - v_{i}$
\begin{eqnarray}
&&\ddot{r}_i = [r_{i-1} ]_+^{3/2}  - 2 [ r_{i} ]_+^{3/2}  -  [r_{i+1} ]_+^{3/2}  +  \kappa (s_i- r_i),  \label{eq:strain1} \\  
&&\ddot{s}_i + \omega^2 s_i = \omega^2 r_i  \label{eq:strain2}
\end{eqnarray}
% \right \}
where $\omega^2 = \omega_0^2-\kappa = \kappa/\nu$. Viewing the right hand side of Eq.~\eqref{eq:strain2} as an inhomogeneity  suggests expressing
the solution in terms of the Green's function of the operator $[\frac{d^2}{dt^2} + \omega^2]$

$$s_i(t) = \omega \int_{-\infty}^t \sin( \omega( t - \tau)) r_i(\tau) d\tau. $$
Substituting this expression into Eq.~\eqref{eq:strain1} yields
\begin{equation}
\ddot{r}_i = [r_{i-1} ]_+^{3/2}  - 2 [ r_{i} ]_+^{3/2}  +  [r_{i+1} ]_+^{3/2}  
+\kappa \left(  \omega \int_{-\infty}^t \sin( \omega( t - \tau)) r_i(\tau) d\tau \right)
-  \kappa  r_i.   \label{eq:strain3}   
\end{equation}
In the absence of the local resonators ($\kappa=0$), it is well known that Eq.~\eqref{eq:strain3} possesses
solitary wave solutions \cite{friesecke94,pego05,stefanov1}. One of the earlier results of Nesterenko relates the amplitude $A$
of the traveling wave to its velocity~\cite{Nesterenko2001}
$$r_i(t) = A \cdot S( A^{1/4} t - i)$$ 
where $S(\xi)$ is the profile of the wave. While there are several analytical approximations of this profile, based on e.g. a long wavelength approximation \cite{Nesterenko2001,ahnert09,yul1}, we use the one
recently developed in~\cite{yul1}, which has the form
\begin{equation} \label{pade}
S(\xi) = \left(\frac{1}{q_0 + q_2 \xi^2 + q_4 \xi^4  + q_6 \xi^6  + q_8 \xi^8 } \right)^2
\end{equation}
where $q_0 \approx 0.8357$,$q_2 \approx 0.3669$, $q_4 \approx 0.0831$, $q_6 \approx 0.0125$ and $q_8 \approx 0.0011$. If we assume that traveling wave form is maintained in the presence of the local resonators,
but has a possibility decaying amplitude, then we have
\begin{equation}
r_i(t) = A_i S( A_i^{1/4} t - i),
\end{equation}
where $S$ is given by Eq.~\eqref{pade}. Substituting this expression into Eq.~\eqref{eq:strain3} yields
\begin{equation}
A_i^{3/2} S''_i = [A_{i-1} S_{i-1} ]_+^{3/2}  - 2 [ A_{i} S_{i}  ]_+^{3/2}  +  [A_{i+1} S_{i+1} ]_+^{3/2}  
+\kappa A_i \left(  \omega \int_{-\infty}^t \sin( \omega( t - \tau))  S( A_i^{1/4} \tau - i) d\tau \right)
-  \kappa  A_i S_i   \label{eq:strain4}   
\end{equation}
where we used the notation $S_i = S(A_i^{1/4} t - i)$.
Since the peak of the traveling wave occurs for $T_i = i A_i^{-1/4}$, and the tail of the solutions asymptotes
monotonically to zero, we can write Eq.~\eqref{eq:strain4} as a discrete map by integrating from $(-\infty,T_i)$
\begin{eqnarray*}
0 =&& A_{i-1}^{3/2} \int_{-\infty}^{T_i} S^{3/2}(A_{i-1}^{1/4} t + i-1) dt  - 2  A_{i}^{3/2}
\int_{-\infty}^{T_i} S^{3/2}(A_{i}^{1/4} t + i) dt  +  A_{i+1}^{3/2}\int_{-\infty}^{T_i} S^{3/2}(A_{i+1}^{1/4} t + i+1) dt  \\
&&+\kappa A_i \left(     \omega  \int_{-\infty}^{T_i}  \int_{-\infty}^t \sin( \omega( t - \tau))  S( A_i^{1/4} \tau - i) d\tau \,dt \right)
-  \kappa A_i \int_{-\infty}^{T_i}  S(A_{i}^{1/4} t + i) dt      
\end{eqnarray*}
where we assumed $A_i>0$ and $S_i>0$. Several of the integrals appearing in this expression
can be evaluated after a change the variable  $t\rightarrow A^{1/4}t +i$. Thus we define
\begin{eqnarray*}
f_1 &=& \int_{-\infty}^{0} S ^{3/2}(t - 1) \, dt \approx 2.5051, \\
f_2 &=& \int_{-\infty}^{0} S ^{3/2}(t) \, dt \approx 1.3215,\\
f_3 &=& \int_{-\infty}^{0} S ^{3/2}(t + 1) \, dt \approx 0.1379,\\
g &=& \int_{-\infty}^{0} S(t) \,  dt \approx 1.2551\\
\end{eqnarray*}
such that
\begin{eqnarray*}
0 =&& A_{i-1}^{5/4} f_1  - 2  A_{i}^{5/4} f_2  +  A_{i+1}^{5/4} f_3  +\kappa A_i \left(     \omega  \int_{-\infty}^{T_i}  \int_{-\infty}^t \sin( \omega( t - \tau))  S( A_i^{1/4} \tau - i) d\tau \,dt \right)
-  \kappa A_i ^{3/4} g.  
\end{eqnarray*}
 \begin{figure}
  \centerline{
\includegraphics[width=.5 \linewidth]{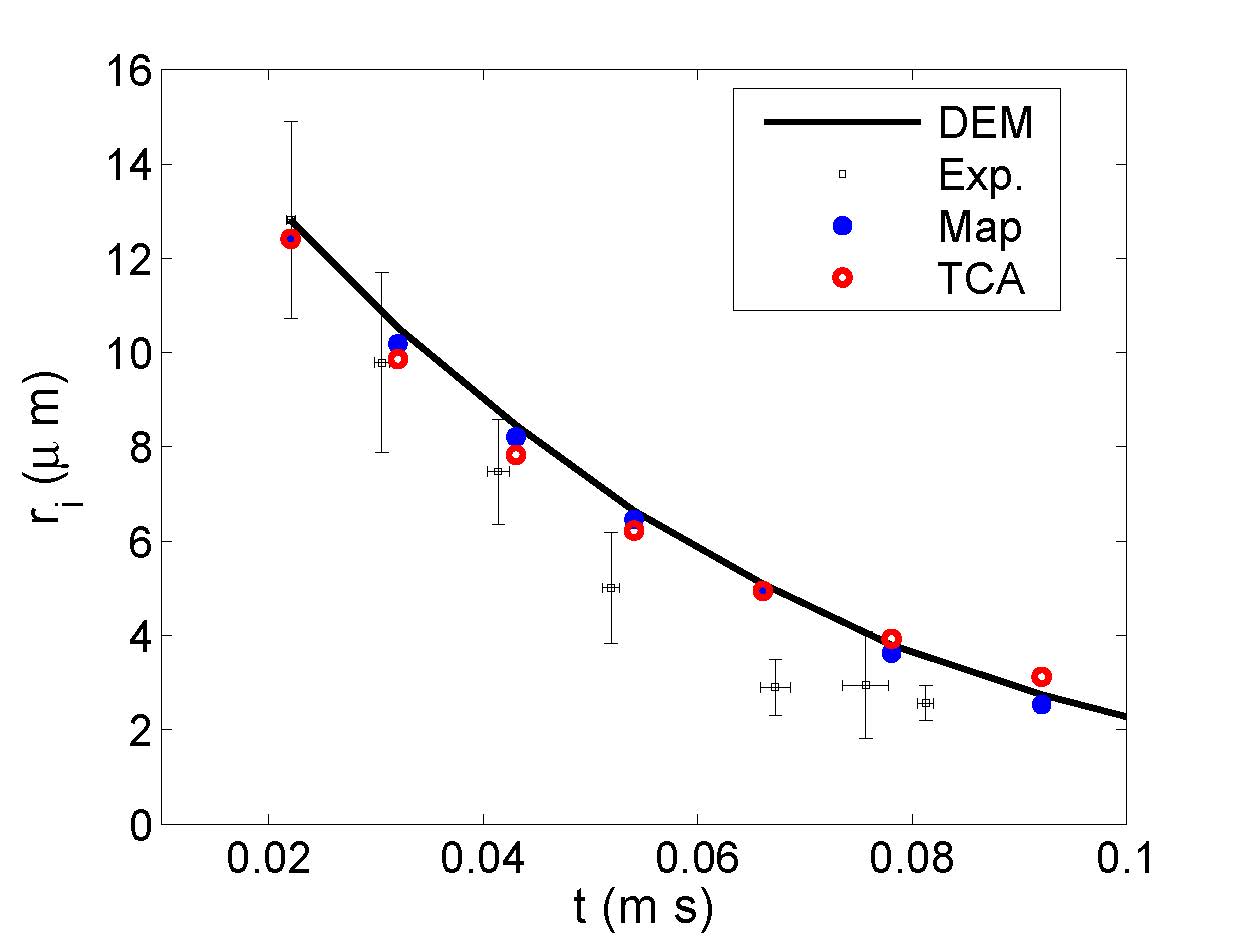}
}
 \caption{Arrival times and corresponding strains $r_i$ of the initial traveling wave front in the $40$ mm rod set up as predicted by the
DEM simulation (black line), the discrete map (blue points) and the TCA (red circles). The squares with error bars
are the experimentally measured values, where the standard deviations were computed using five experimental runs.  }
 \label{fig:arrival}
\end{figure}
Assuming that local resonators create small perturbations of size $\epsilon$
such that $A_{i-1} = A_i + \mathcal{O}(\epsilon)$ finally yields a one dimensional discrete map
%$A^{5/4}_{i+1} \approx A^{5/4}_{i}$ and that $A_i \approx A_{i-1}$, which finally yields,
\begin{equation} \label{eq:map}
A_{i+1} = \left(\frac{A_{i}^{5/4} f_1 + \kappa A_i \left(     \omega  \int_{-\infty}^{T_i}  \int_{-\infty}^t \sin( \omega( t - \tau))  S( A_{i}^{1/4} \tau - i) d\tau \,dt \right)-   \kappa A_i ^{3/4} g}{2f_2 - f_3}\right)^{4/5}.
\end{equation}
By construction, the arrival time at the $i$th lattice site will be $T_i = i A_i^{-1/4} + \phi_0$, where $\phi_0$ is used to calibrate
the position of the first peak. To check the validity of this approximation, we compare the approximate decay of the peaks and corresponding arrival times to 
the DEM simulations. In Fig.~\ref{fig:arrival} we consider the parameter values determined by the optimization procedure described in Table I
for a rod length of $40$ mm. In this case, the arrival time and strain value $r$ of the initial traveling wave front compare favorably among the experiment, full DEM calculation, and map based on Eq.~\eqref{eq:map}.

\begin{figure}
  \centerline{
\includegraphics[width=.4 \linewidth]{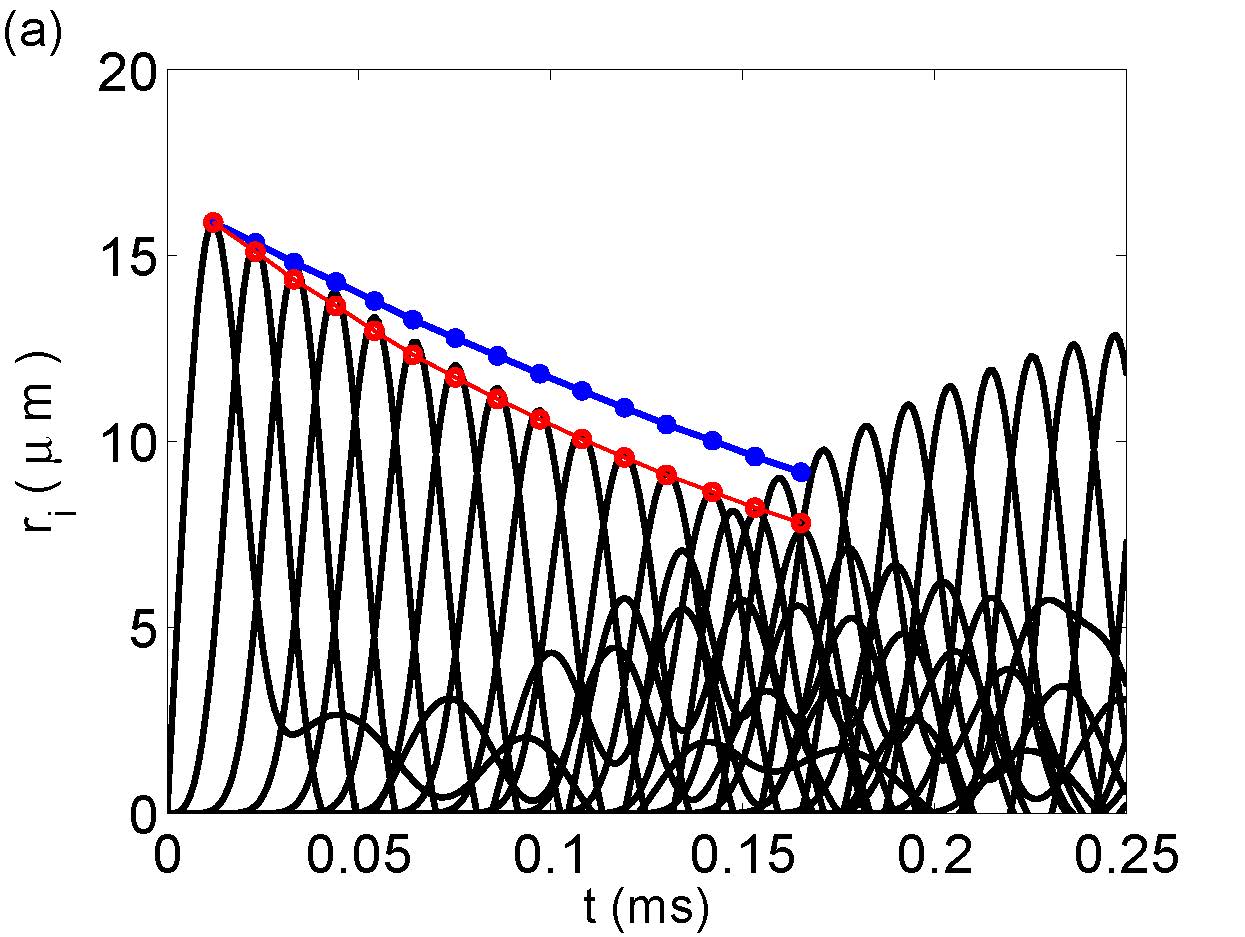}
\includegraphics[width=.4 \linewidth]{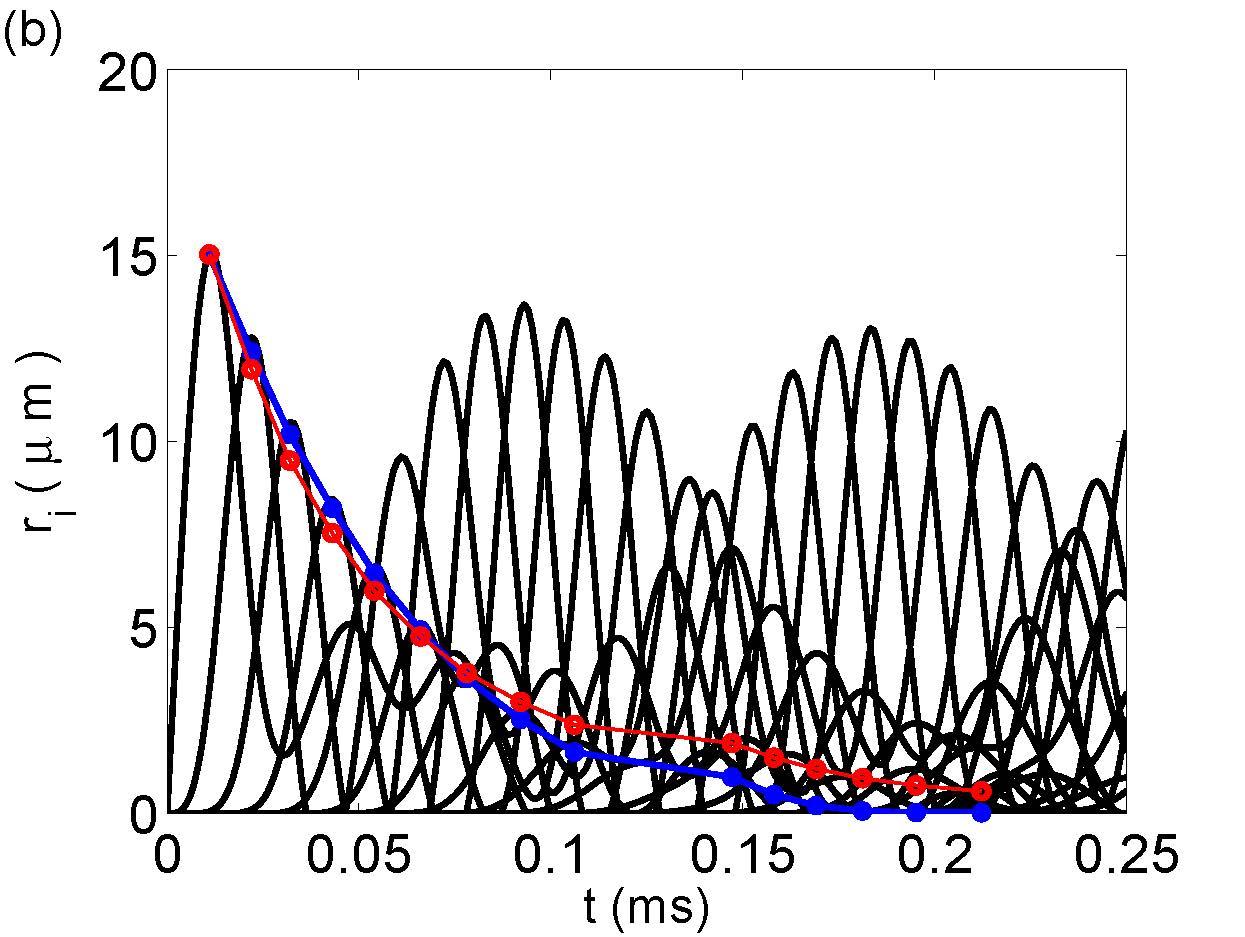}
}
\vspace{.1cm}
  \centerline{
\includegraphics[width=.4 \linewidth]{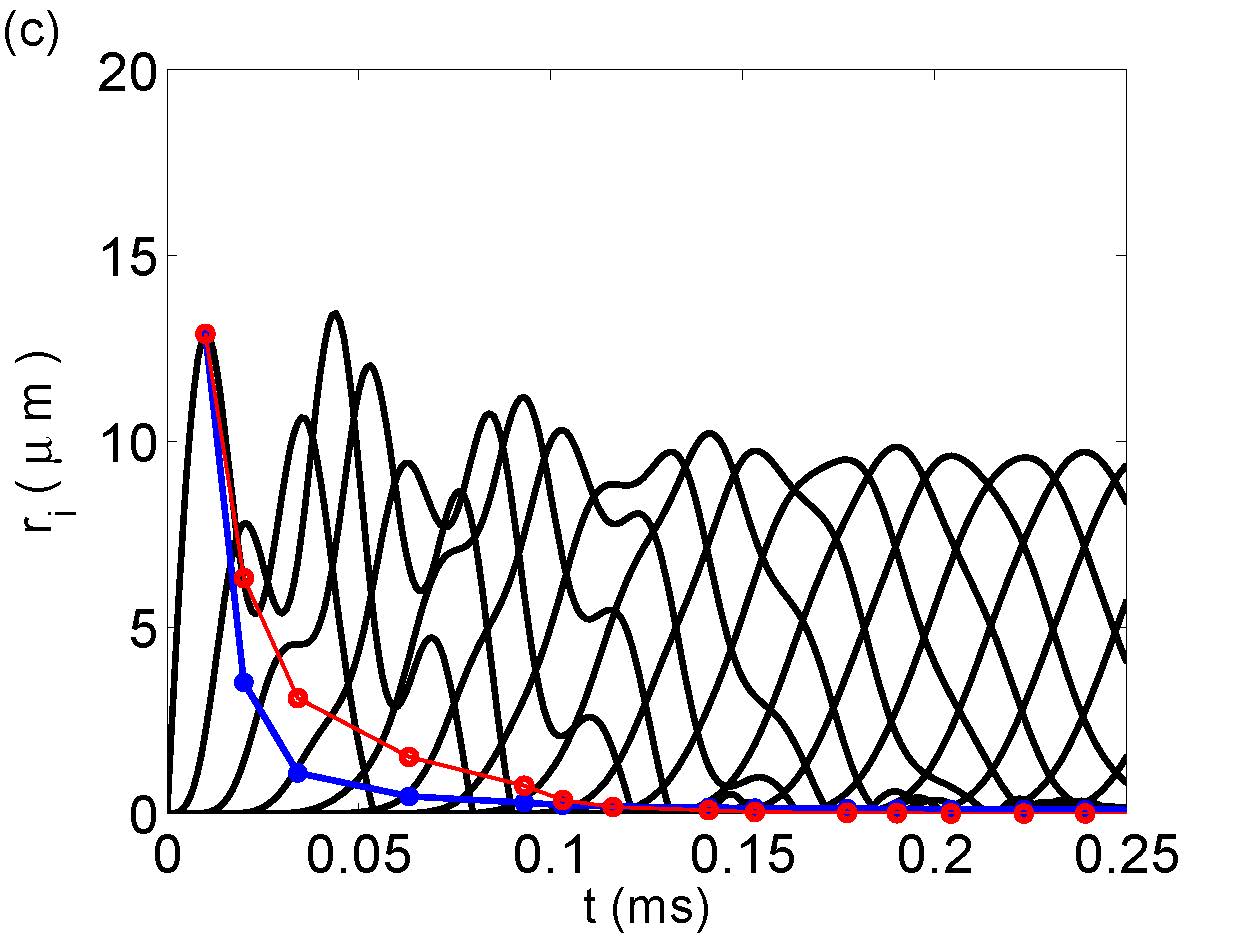}
\includegraphics[width=.4 \linewidth]{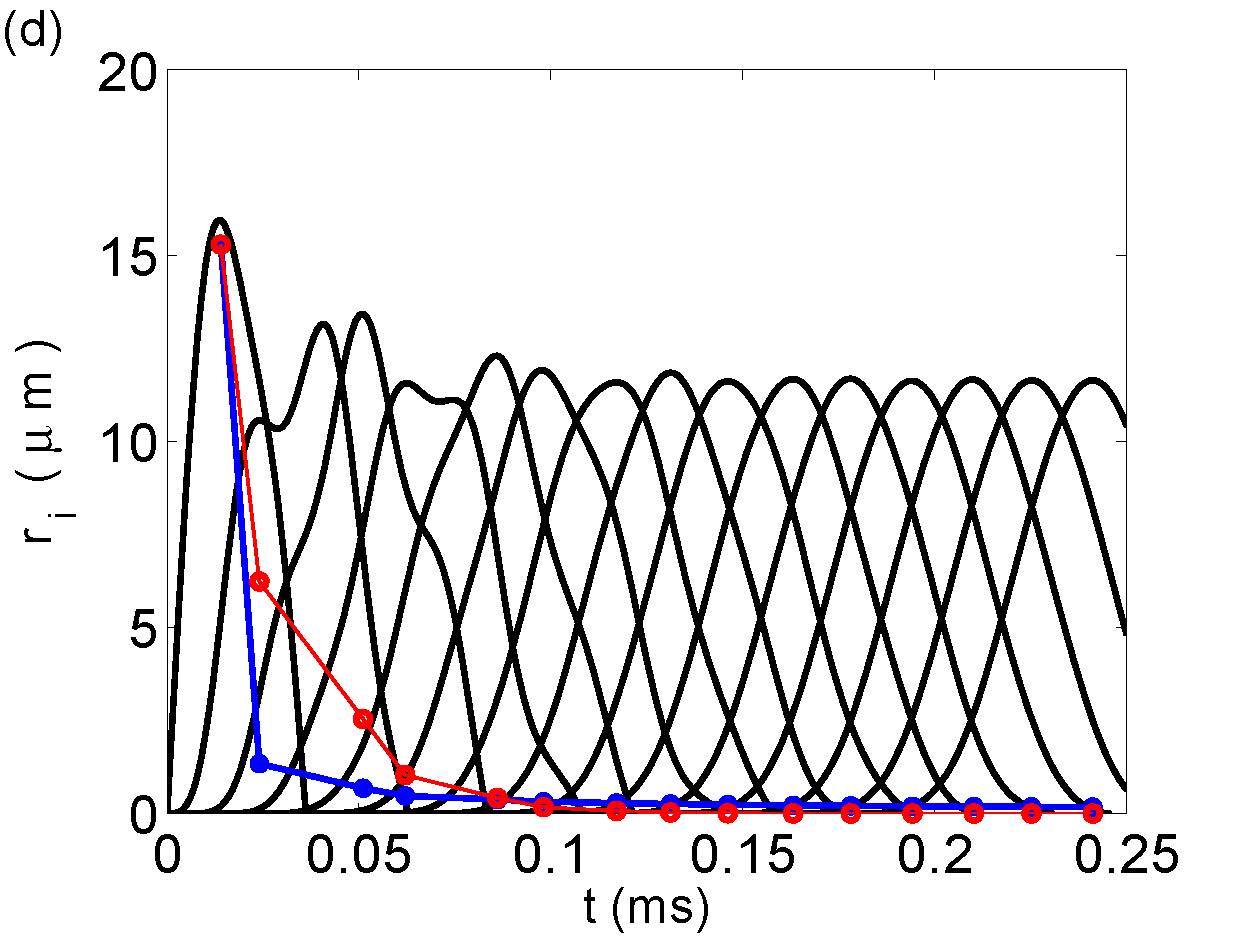}
}
 \caption{Plot of the strain $r_i$ for beads $i=1,\dots,15$ from the full DEM simulation (black lines) and the decay of the peaks as predicted by the map approach
(blue points) and the TCA (red circles).  All parameter values correspond to those determined by the optimization procedure described in Table I
with the exception of the
resonant frequency $k_1$ (kN/M), which has the value (a) $k_1 = 788$, (b) $k_1 = 3961$, (c) $k_1 = 15760$, and (d) $k_1 = 23641$.
The resonant frequency of panel (b), $k_1 = 3961$, corresponds to the value in Table I for the 40mm rod.
Due to the immediate effect of the secondary waves in the higher frequency cases shown in panels (c-d), both the map approach and TCA are inaccurate.
In panel (d) one can see how the core part of the nanopteronic forms. The tail part, however, cannot be discerned at this resolution. }
 \label{fig:decay}
\end{figure}

In the main text, three types of behavior are observed based on rod length, namely (i) the spontaneous formation (and steady propagation)
of the nanopteron %SVN 
(found in the 20mm rod), (ii) the potential breathing of the traveling solitary waves (in the 40mm rod)
or (iii) the decay of the solitary waves (in the 80mm rod) without relying on internal damping. This is due to the fact that each rod length yields different
values of the resonant frequency. Thus, varying the resonant frequency will
allow one to observe behavior ranging from modulated waves (i.e., breathing) to sonic-vacuum-nanoptera. For example, 
Fig.~\ref{fig:decay} compares simulations of the DEM model and the map (Eq.~\eqref{eq:map})
for other values of the local resonant frequency $\kappa$, while keeping the other parameter values fixed. The modulation effect, which is due to the presence of secondary waves, is not captured by the map approach, 
which only takes into account the effects of the initial wave. Smaller values of the
parameter $\kappa$ correspond to slower decay of the initial traveling wave front, which is captured well by the map approach. Likewise, as $\kappa$ is increased, the decay
rate increases. However, in the full DEM model, the secondary wave effects have an immediate impact, which the map approach
fails to predict, see Fig~\ref{fig:decay}(c) and (d). In  panel (d), one sees the development of the nanopteron.
Note that the role of the secondary wave in numerical 
observations was also discussed in~\cite{staros2}.
In Figs.~\ref{fig:arrival} and \ref{fig:decay} the predictions based on a ternary collision approximation (TCA) are
also shown, which is detailed in the following section.

\subsection{An Extension of the BCA: A Ternary Collision Approximation}

The main purpose of the binary collision approximation (BCA) is to offer an 
approximation of the solitary wave velocity, or rather, the arrival time of the pulse for a given bead. %\cite{Lindenberg}.
The idea is to solve a simplified set of equations over a time scale where the dominant effects are between two adjacent beads
only. In some cases, the resulting simplified equations can be solved exactly, the solution of which can be used
to initialize the following step. Stringing the approximations together yields an approximation over the desired time span.

Employing the same procedure for the DEM model, Eqs. (1) and (2) in the main manuscript leads to a ternary collision approximation
(since two primary masses and a local resonator are now involved).
The simplified equations in the rescaled variables have the form
\begin{eqnarray*}
 \ddot{u}_0 &=&  [ u_0 - u_1 ]_+^{3/2} -  \kappa ( u_0 - v ), \\ 
 \ddot{u}_1 &=&  [ u_0 - u_1 ]_+^{3/2}, \\
 \ddot{v} &=&   \omega^2 ( u_0 - v ). 
\end{eqnarray*}
where  $u_0$ is the displacement of the impacted bead, $u_1$ is the displacement of the bead adjacent to it, and $v$ is the displacement of the local resonator
of the impacted bead.  
Now define $r = u_0 - u_1$ and $s= u_0 - v$. We have then
\begin{equation} \label{eq:TCA_strain}
\begin{array}{ll}
 \ddot{r} =& -  2 [r]_+^{3/2} - \kappa  s, \\ [2ex]
 \ddot{s} =&  - [ r ]_+^{3/2} -  \left( \kappa + \omega^2 \right) s.  
\end{array} 
\end{equation}
The goal is to use these equations to obtain an approximation for the arrival time of the pulse.
Recall that the arrival time at the $i$th site corresponds to the time where
$r_i(t)$ obtains its first local maximum, or equivalently when the velocities of the adjacent beads
become equal, since by definition $r_i = u_{i-1} - u_i$. 
For the first iteration, we use the initial value $r(0)=s(0)=0$ and $\dot{r}(0)=\dot{s}(0)=V_0$,
where $V_0$ is the velocity of the striker bead.
Upon solving system~\eqref{eq:TCA_strain}, we define the first arrival time $t_1$
such that $\dot{r}(t_1) = 0$. For the subsequent iterations, we use the initial value
$r(0)=s(0)=0$ and $\dot{r}(0)=\dot{s}(0)=V_i$. In the absence of the local resonator,
the traveling wave does not decay, and hence $V_i = V_0$. We assume however, that resonator
will indeed cause the peak to decay. We estimate the decay based on the first iteration.
Namely, we define
$$ V_i = V_0 \, a^{i-1},  \qquad a = \frac{\dot{r}(t_1;\kappa)}{\dot{r}(t_1;0)}$$
where $r(t_1;0)$ is the solution of~\eqref{eq:TCA_strain} at time $t=t_1$ with the resonator parameter $\kappa=0$
and $r(t_1;\kappa)$ is the solution of~\eqref{eq:TCA_strain}  at time $t=t_1$ with the resonator parameter $\kappa\neq0$.
In other words, $a$ is the decay between the first and second peak, and we assume
the decay rate remains constant. 
Since the TCA is an iterative procedure, 
the $i$th arrival time $t_i$ is based on an initial time of $t=0$.
Thus, the actual arrival time at site $i$ is  $T_i = \sum_n^i t_n$. 
The pulse arrival times based on the TCA using the parameter values corresponding to the $40$ mm rod 
are shown as the red circles in Fig.~\ref{fig:arrival}, where we found $a\approx 0.8$. This method also compares favorably to the experiment, full DEM simulation
and the map approach. To estimate the amplitude of peak $i$,
we used the formula $A_0 a^{i}$, where $A_0$ is the 
amplitude of the first peak computed from the DEM. These predictions are shown as red circles in
Fig.~\ref{fig:decay}.

Here, the TCA reduced an infinite dimensional
system of ODEs into a system of two second order equations (see e.g. Eq.~\eqref{eq:TCA_strain}). However, the simplified system does not lend itself
to a closed form analytical solution, and thus we resorted to numerical simulations of  Eq.~\eqref{eq:TCA_strain}.
Nonetheless, the favorable agreement between the TCA and full DEM simulations demonstrates that the predominant
features of the initial traveling wave front at a given time are adequately
 captured by the simplified model bearing 
two primary masses and a local resonator.